\documentclass{article}       
\usepackage{graphicx}
 \newcommand*\diff{\mathop{}\!\mathrm{d}}
 \usepackage{physics}
 \usepackage{float}
\begin{document}

\title{Single Photon Two-Level Atom Interactions in 1-D Dielectric Waveguide: Quantum Mechanical Formalism and Applications }


\author{Fatih Din\c{c} 
\and 
        \.{I}lke Ercan\thanks{The authors are with the Electrical and Electronics Engineering Department in Bogazici University.}
       }


\date{Version: \today} 
\maketitle

\begin{abstract}
In this paper, we propose an effective model including a macroscopic Hamiltonian to describe the interactions between a two-level atom and scattered light in a 1-D dielectric waveguide. The proposed formalism allows us to incorporate the effect of changing optical media inside the continuum while demonstrating a non-classical derivation of Fresnel Law. We obtain the transport characteristics of the two-level system, explore its high-Q bandreject filter property and discuss the implications of radiative and non-radiative dissipation. In addition, we apply our formalism to a modified Fabry-P\'{e}rot interferometer and show the variation in its spontaneous emission characteristics with changing interferometer length. Finally, we conclude with further remarks on the link between the waveguide and cavity quantum electrodynamics.
\end{abstract}



\section{INTRODUCTION}

Two-level atom coupled to a 1-D continuum has been a major topic of interest due to its potential to predict a mirror behaviour of the atom for certain wavelengths of light. Recent studies reveal that spontaneous emission in such systems can lead to various properties that have been investigated in \cite{fan,zang,fandissipation,Law,fan-2,fan-3}. The position space Hamiltonian approach, as suggested in \cite{fan}, can be used for various calculations such as single atom in waveguides, Fano interference, a linear chain of atoms in waveguide or an atom coupled to two waveguides.  However, to the best of our knowledge, current literature does not present a quantum mechanical formalism that allows incorporating the variation of the optical medium inside the continuum. This is particularly important since single photon switches using waveguide-cavity and their experimental realizations are drawing attention \cite{ilkehoca,ilkehoca-2,ilkehoca-5,ilkehoca-6}. Understanding and exploiting Kerr nonlinearities are also important as the presence  of an atom in the cavity may present an opportunity to use such systems in quantum computation \cite{ilkehoca-3,ilkehoca-4}.  \par
The existing treatments of two-level atoms (two-level system (TLS)) put particular emphasis on  its transport characteristics for various situations as in \cite{kocabas,kocabas2,rephaeli}. \cite{shi,cavity} describe the interaction between a TLS inside a cavity and a waveguide.  The routing effect of chiral coupling for the TLS has been investigated in \cite{chiral}. The system consisting of multiple resonator arrays are presented in \cite{resonator}. In all of these listed articles, as well is in existing literature, the waveguide is assumed to be consisting of a material which has constant refractive index everywhere.

The main contribution of our work to the literature lies in the formalism we introduce to incorporate the effect of optical media inside waveguides. Our approach is an effective preliminary attempt to understand the quantum nature of the dielectric media using a position space Hamiltonian. Developing such formalism plays a crucial role in enabling comparison between theoretical derivations and experimental results. Certain classical approaches allow incorporation of an optical medium through classical wave equation inside waveguides \cite{photonics}, however, treating interactions with two-level atoms  require deriving purely quantum mechanical formalism. The modified Fabry-P\'{e}rot interferometer we propose in this article is one such example, where the presence of the interferometer may change the life-time of the TLS. This property might be of immense importance for future quantum computing and switching applications of the TLS. Moreover, our work allows one to glance at a part of what might be the true quantum nature of the Fresnel Law. Due to approximate nature of our derivations, we omit arriving at decisive conclusions regarding its nature. Nonetheless, our calculations hint that the Fresnel Law might have a different perspective in quantum mechanics than in classical mechanics. In supplementary materials, we show the correspondence between our work and the literature, especially  \cite{fan,Law} for dielectric waveguides with constant refractive index everywhere.

The organization of this paper is as follows: In section 2, we begin by laying out the foundations of the proposed formalism,  derive the position space Hamiltonian and incorporate the presence of an optical medium as well as changes in refractive index in the framework. In section 3, we illustrate our formalism  through application to a TLS inside a 1-D dielectric waveguide. We also discuss applications to high-Q bandreject filters and effect of dissipation. In Section 4, we propose a modified Fabry-P\'{e}rot interferometer and show how its spontaneous emission characteristics differ from the two-level system. Furthermore, we comment on a link between waveguide and cavity QED. In the last section, we conclude with final remarks and comments on future directions.

\section{Formalism}

\subsection{Derivation of The Hamiltonian}
In this section, we layout the theoretical foundation of the new formalism we propose using position space Hamiltonian.\footnotemark \footnotetext{In this article, operators will be expressed using hats (such as $\hat{f}$) and unit-vectors will be expressed using the inverse-hats (such as $\check{k}$). This representation is adapted from Dr. Austin J Hedeman's course notes at University of California Berkeley.} The departure point of this formalism is the quantum harmonic oscillator behavior of light and field creation/annihilation operators in position space. The distinction between this formalism and that of \cite{fan} and \cite{Law} is that we use the particle creation operator $\phi^\dag(x)$ (defined in \cite{qft-operator}) without dividing the Hilbert space into subspaces, such as the ones containing right-moving and left-moving particles. \par

The Dicke--Hamiltonian \cite{dicke-hamiltonian} describes the interaction between a two-level atom and light. It can be transformed, using periodic boundary conditions \cite{periodic-boundary}, into continuous k-space \cite{continuous-k-space}. The transformation from continuous k-space to position space is then followed by defining the position space components of continuous k-space operators. The transformed position space Hamiltonian is
\begin{eqnarray}
  \label{eqhamiltonian:2}
		\hat H &= -i \hbar v_g \hat f \cdot \int_{-\infty}^\infty \diff x   \phi^\dag(x) \grad \phi(x) + E_e a_e^\dag a_e  + E_g a_g^\dag a_g  \nonumber
		 \\ &+V'  \int_{-\infty}^{\infty}\diff x \delta(x) 
		 \left( \phi^\dag(x) S_-  + \phi(x) S_+ \right), 
\end{eqnarray}
where $v_g$ represents the group velocity of the photon inside the waveguide, $\ket 0$ the vacuum state, $E_e$  and $E_g$ the energy of the excited and ground state of the atom, respectively, with $\Omega=E_e-E_g$. $V'$ is the coupling coefficient between the photon and the two-level atom, $S_-=a^\dag_g a_e$ ($S_+=a^\dag_e a_g$) is the atomic de-excitation (excitation) operator -- i.e. $a_{e/g}=\ket{0}\bra{e/g}$ ($a^\dag_{e/g}=\ket{e/g}\bra{0}$) is the destruction (creation) operator for the excited/ground state of the atom. The unit k-vector operator $\hat f$ is defined as $\hat f \ket{k} = \check x sgn(k) \ket{k}$. The equivalence of our approach and that of \cite{fan})is illustrated in Appendix~\ref{proofofconcept}. It is important to note that the delta-scattering potential $V \delta(x)$ is highly peaked at the origin and approximately nonexistent at other points of space, which is inline with our assumption that the interaction between light and TLS is weak \cite{UVcutoff}.\par

\subsection{The position space Hamiltonian of Light Inside Optical Media}
Now, we shall focus on the effective description of the interactions between light and atoms inside the dielectric media. To achieve this, one can start by transforming the macroscopic discrete-k-space Hamiltonian, which is given by
\begin{eqnarray}
  \hat H_{mac} = \sum_k \hbar w_k \hat a_k^\dag \hat a_k, \nonumber
\end{eqnarray}
where $a_k$ ($a_k^\dag$) is the annihilation (creation) operator for a photon inside a linear optical medium. Here, the assumption of linearity is crucial, as we seek to describe the interactions between photon and the optical medium through a macroscopic Hamiltonian in a simple manner. In this model, we perform second quantization by interchanging $\epsilon_0 \iff \epsilon$ and working with the collective electric field $\vec{E}(\epsilon)$.  Afterwards, we transform this macroscopic (effective) Hamiltonian, which is obtained through quantization of $\vec E(\epsilon)$, from discrete k-space to continuous k-space.\footnote{A more detailed discussion for the derivation of the macroscopic Hamiltonian can be found in Chapter 7 of \cite{dutra2005cavity}.} For the scope of this paper, we will linearize the coupling constant (and the macroscopic dielectric field) around $E_k=\Omega$ as a first order approximation. From now on, we shall refer to the macroscopic Hamiltonian $\hat H_{mac}$ simply as the Hamiltonian $\hat H$.

Assuming that the photon is inside a 1-D optical medium with refraction index $n$ and quantization length $L_q \to \infty$, one can write the continuous k-space Hamiltonian as
\begin{eqnarray}
  \hat H = \int_{-\infty}^{\infty} \text{d}\bar{k} \hbar w_k  a^\dag(k) a(k),
\end{eqnarray}
where $\bar{k}=k_0n$ with $k_0$ representing the wave-vector of photon in the absence of optical medium and $a^\dag(k)$/ $a(k)$ are, respectively, the creation and annihilation operators for a photon with frequency $w_k$ and with a wave-vector $k_n=k_0n$ in the continuous k-space. It is important to note that we use the periodic boundary condition described in \cite{periodic-boundary} ($\Delta \bar{k}  = 2\pi/L \nonumber$).
In this definition, the effect of the linear dielectric medium is manifested both as lumped inside the creation and annihilation operators, and through the periodic boundary conditions. Keeping the latter effect in mind, we can define the creation, $\phi^{\dag(n)}(x)$, and annihilation, $\phi^{(n)}(x)$, operators inside an optical medium with a refractive index $n$ as
\begin{equation}  \label{eq23}
\begin{split}
\phi^{(n)}(x) &= \int_{-\infty}^\infty \diff \bar{k}  \bra{x}a^\dag(k)\ket{0} a(k),\\
\phi^{\dag(n)}(x) &= \int_{-\infty}^\infty \diff \bar{k}  \bra{0}a(k)\ket{x} a^\dag(k),
\end{split}
\end{equation} 
where $\ket{0}$ is the vacuum-state for the photon at  position $x$. We note that this definition implies the merging of different mediums at finitely many optical boundaries. If the photon is far away the boundary, the field operators are the Fourier transform of k-space operators. However, the definition of the field operators in the optical boundaries are arbitrarily chosen such that the symmetry for scattering of light from left and right is not broken, e.g. there is a reference refractive index at the boundary which is the average refractive index of the both sides. We shall see in later sections that this idea hints at the derivation of the Fresnel Law from a quantum mechanical perspective.

The definition given in (\ref{eq23}) leads to an important relation for optical media with two different refractive indices $n$ and $n'$
\begin{subequations} \label{eq24}
\begin{align}
\phi^{(n)}(x) &= \frac{n}{n'} \phi^{(n')}(x), \\
\phi^{\dag(n)}(x) &= \frac{n}{n'} \phi^{\dag(n')}(x).
\end{align}
\end{subequations}

This relation constitutes the foundation of our analysis as it allows us to incorporate the optical medium change at the optical boundary. With this new formalism, it is possible to include effects of the transition between different optical media. Keeping this in mind, we can start exploring the properties of light-two level atom interactions.

\section{Applications}
\subsection{Scattering of Light from 1-D Optical Boundary} 

In this section, we consider a photon inside a waveguide, which consists of two different material with different dielectric constants, $n_1$ and $n_2$, with the goal to obtain Fresnel coefficients for normal incidence. In this approach, a reference refractive index $n_r$ is taken for the boundary at $x=0$, yet the results are independent of this reference. The setup is illustrated in Fig. 1.

\begin{figure}
\centering
\includegraphics[width=8cm]{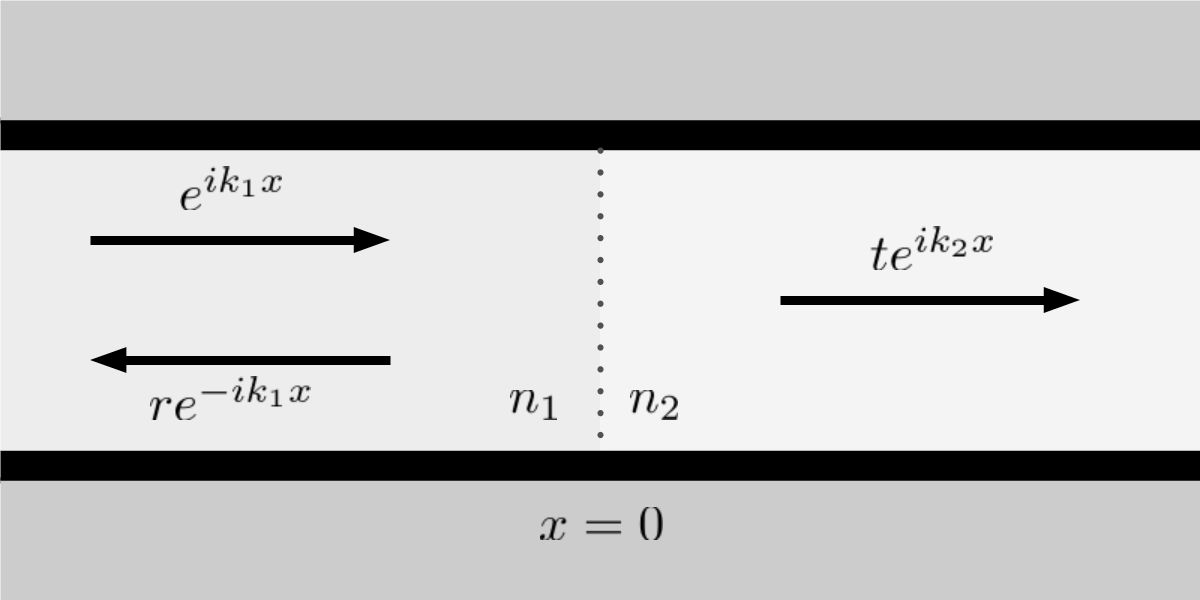} 
\caption{A waveguide of different optical media  on the left (dark gray) and right (light gray) with a boundary at $x=0$. The optical medium to the left of atom has refractive index $n_1$ and one on the right has refractive index $n_2$. } \label{figfresnel}
\end{figure}

The Hamiltonian for this setup is given by
\begin{equation}
\begin{split}
 \hat H &=- i \hbar \frac{v_g}{n_1} \hat f \cdot \int_{-\infty}^{0} \text{d} x \,  \phi^{\dag(n_1)}(x) \grad \phi^{(n_1)}(x)  \\ &-i \hbar \frac{v_g}{n_2} \hat f \cdot \int_{0}^{\infty} \text{d} x \,\phi^{\dag(n_2)}(x) \grad \phi^{(n_2)}(x),   
\end{split}
\end{equation}
where $v_g/n_1$ and $v_g/n_2$ represent the group velocity of photon inside the two distinct regions of waveguide.

The eigenstate of this Hamiltonian for a photon with frequency $w_k$ is
\begin{equation}
\begin{split} 
	\ket{E_k}&=\int_{-\infty}^{0^-} \diff x \left(e^{ik_0n_1x} u_{R1}  (x)+ e^{-ik_0n_1x}u_{L1}(x)\right) \phi^{\dag(n_1)}(x)  \ket{0} \\ &+  \int_{0^+}^{\infty} \diff x \left(e^{ik_0n_2x} u_{R2}(x)+ e^{-ik_0n_2x}u_{L2} (x')\right)\phi^{\dag(n_2)}(x)  \ket{0} \\
	&+ \int_{0^-}^{0^+} \diff x \Big( e^{ik_0n_rx}(u_{R1}(x)+u_{R2}(x))  + e^{-ik_0n_rx}(u_{L1} (x) + u_{L2}(x))\Big) \phi^{\dag(n_r)}(x)\ket{0},
	\end{split}
\end{equation}
where the boundary is taken to be diminishing.  Here, the complex coefficients $u_{L1}(x)$,  $u_{L2}(x)$, $u_{R1}(x)$  and $u_{R2}(x)$ correspond to the photon moving towards either left, L, or right, R in medium 1 or 2, respectively as shown in Fig. \ref{figfresnel}.

Solving the eigenvalue equation $\hat H \ket{E_k}=E_k \ket{E_k}$, with the realization of the relation given in (\ref{eq24}) and taking the boundary width as diminishing (which will be justified later on), one obtains the following Fresnel coefficients \cite{fresnel} 
\begin{align} \label{fresnel}
t= \frac{2n_1}{n_1+n_2}, \quad \quad r=\frac{n_1-n_2}{n_1+n_2}.
\end{align} 
A more rigorous analysis requires one to understand the shape of the scattering potential, created by the boundary, and the Fourier components (k-components) of the wave function.\footnote{It is important to note that in this calculation, we neglect the contribution from some interference terms which are not individually zero. Hence, our result remains as an approximation in the region where the contribution from these interference terms can be neglected, i.e. $\frac{|n_1-n_2|}{n_1+n_2} << 1$.} We believe that a more rigorous analysis can shed light on the nature of the Fresnel Law. For the rest of this article, we will assume that this interference terms can be neglected in alignment with the envelope approximation. A generalization of our treatment in this section can be applied to the incidence of 2-D s-polarized light. Employing a very similar approach, one can show that the Fresnel law and the Snell's law can be achieved in the classical limits for this specific case as well.

We note one final remark on the effect of the diminishing reference boundary. If the operator $U(n_1 \to n_r)$ describes the transition of light from the medium one to the reference boundary, then the effect of the reference medium can be represented by the operation $M=U(n_1 \to n_r)P U(n_r \to n_2)$, where $P$ represents the propagation inside the reference boundary. It can be shown through matrix algebra that as the width of the reference boundary diminishes, $P \to I$, it has no effect on the transport characteristics of the system, e.g. $M \to U(n_1 \to n_2)$, justifying our use of diminishing reference boundary in the energy eigenstate.

\subsection{Light--Two-Level Atom Interactions}
\subsubsection{Preliminaries}
\begin{figure}
\centering
\includegraphics[width=8cm]{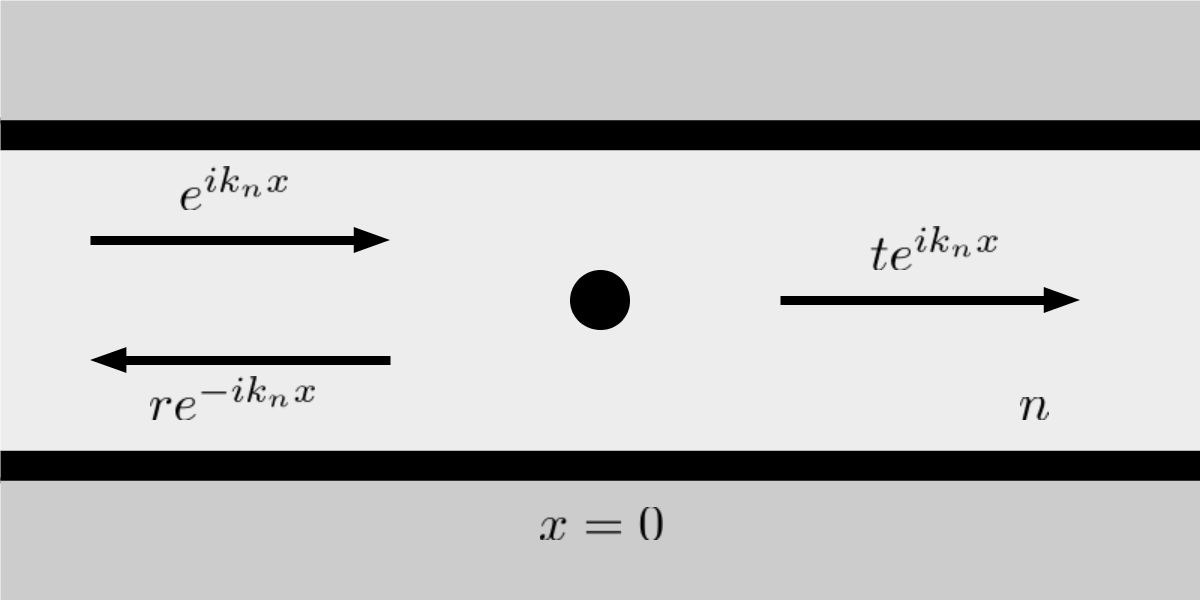} 
\caption{The two-level atom inside a waveguide with a refractive index $n$. } \label{figmain}
\end{figure}

 In this section, we derive the relations governing the coupling between a photon and two-level atom in an optical medium by using a quantum mechanical treatment. Fig. \ref{figmain} depicts the photon scattering for a system where the two-level atom is situated in a waveguide with a refractive index $n$.

 The Hamiltonian for this system is given as\footnote{This representation of the interaction Hamiltonian differs from (1) in the two following ways: The field operators $\phi(x)$ are replaced with $\phi^{(n)}(x)$ and the group velocity $v_g$ of the photons is replaced with $v_g/n$ to incorporate the effect of dielectric medium, $v_g$ corresponding to the group velocity in vacuum.}
 \begin{eqnarray}
 \hat H&=- i \hbar \frac{v_g}{n} \hat f \cdot \int_{-\infty}^{\infty} \text{d} x \,  \phi^{\dag(n)}(x) \grad \phi^{(n)}(x)+ \Omega a_e^\dag a_e  \nonumber \\
 		&+ V_n  \int_{-\infty}^{\infty}\diff x \delta(x) (  \phi^{\dag(n)}(x) S_-  + \phi^{(n)}(x) S_+  ), \label{Hamiltonianeq}
 \end{eqnarray} 
where $V_n$ is the coupling coefficient between the photon and the atom situated in an optical medium with a refractive index $n$. We restrict $V_n\geq V$ in alignment with the increased decay rate inside the dielectric media. A more general discussion on the radiative decay rate modification inside dielectric medium can be found in  \cite{qopticsdielectric}. For the purpose of our formalism, we shall stress that the coupling coefficient $V_n$ is a highly linearized parameter, which allows an effective, even ad-hoc, exploration of the matter-light interactions inside dielectric medium.

A scattering eigenstate $\ket{E_k}$ of this Hamiltonian for a photon, incident from left, with frequency $w_k$ is given by 
\begin{equation}\label{eq:eigenstate}
\begin{split} 
	\ket{E_k}&=\int_{-\infty}^{\infty} \diff x \left(e^{ik_n x} u_{R}  (x)+ e^{-ik_nx}u_{L}(x)\right) \phi^{\dag(n)}(x)  \ket{0,-} + e_k\ket{0,+}.
\end{split}
\end{equation}
Here, the complex coefficients $u_{L1}(x)$,  $u_{L2}(x)$, $u_{R1}(x)$  and $u_{R2}(x)$ correspond to the photon moving towards either left, L, or right, R in medium 1 or 2, respectively as shown in Fig. \ref{figmain}.

Solving the eigenvalue equation $\hat H \ket{E_k}=E_k \ket{E_k}$, one obtains the following set of equations
\begin{subequations}
\begin{align}  
		& i \hbar \frac{v_g}{n} \left( \frac{\partial u_{L1}(x)}{\partial x} e^{-ik_n x} - \frac{\partial u_{R1}(x)}{\partial x} e^{ik_nx} \right)+  V_n \delta(x)  e_k \notag \\ 
		&+  i \hbar \frac{v_g}{n} \left(\frac{\partial u_{L2}(x)}{\partial x} e^{-ik_nx} - \frac{\partial u_{R2}(x)}{\partial x} e^{ik_nx} \right) = 0, \\
		& V_n t + (\Omega -E_k) e_k=0.
\end{align} 
\end{subequations}

The solution of these equations are obtained while taking the wave continuity condition at $x=0$ into account
\begin{eqnarray}
 r &= i \sin b e^{ib},\nonumber \\
  t &=  \cos b e^{ib}, \label{experiment} \\
     e_k&=-\frac{v_g}{nV_n} \sin b e^{ib},\nonumber
\end{eqnarray}
where the phase shift is given as $b=\arctan\{V_n^2 n/[ v_g(\Omega-E_k)]\}$. Note that, in (\ref{experiment}), we set $\hbar=1$ in order to use energy and corresponding frequency values interchangeably. \par

\begin{figure*}
\centering
\includegraphics[width=8cm]{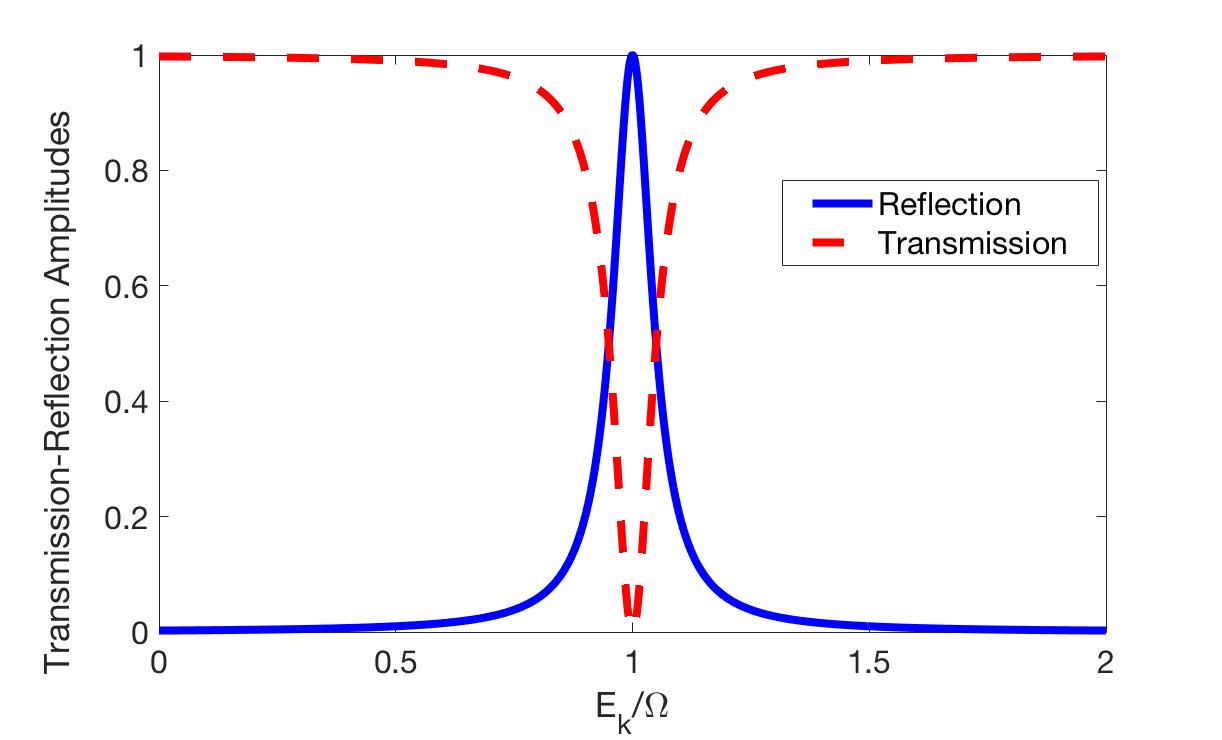} 
\caption{Transmission (dotted) and reflection (solide) amplitudes for $J_n = 0.05\Omega$ vs $E_k/\Omega$.  } \label{figtrans}
\end{figure*}

The transfer matrix for a two-level atom with two different optical media at the optical boundary is defined as
\begin{eqnarray}
\begin{pmatrix}
E_L^{(+)}  \\
E_L^{(-)} 
\end{pmatrix}
=
S_{[2\times2]}
\begin{pmatrix}
E_R^{(+)}  \\
E_R^{(-)}
\end{pmatrix},	\nonumber
	\end{eqnarray}
where $E_{L/R}^{(\pm)}$ is the amplitude of electric field with positive/negative wave-vector on the left/right of the two level atom.
After some algebra, the transfer matrix is given by 
\begin{equation}  \label{eq:smatrix}
S_{2\times2}^{\text{two-level system}}=
\begin{bmatrix}
1 - i \frac{J_n}{(\Omega - E_k)} & - i \frac{J_n}{(\Omega - E_k)} \\
  i \frac{J_n}{(\Omega - E_k)} & 1 + i \frac{J_n}{(\Omega - E_k)}
\end{bmatrix},
\end{equation}  
where we define $J_n= \frac{V_n^2n}{ v_g}$ representing the coupling energy in dielectric. Here, it is important to emphasize that since $J_n$ has the dimensions of energy it can be obtained experimentally.

Moreover, the reflection amplitude is a Lorentzian, which can be written as
\begin{eqnarray}
R = |r|^2 =  \sin^2 b =\frac{J_n^2}{J_n^2 +(\Omega-E_k)^2}. 
\end{eqnarray}

It can be easily shown that $2J_0$ is the bandwidth for the ultra-narrow filter in the presence of the vacuum. In the presence of the dielectric medium, the bandwidth of the resulting Lorentzian broadens and becomes $2J_n \geq 2J_0$. In Fig. \ref{figtrans}, the transmission and reflection amplitudes ($T=|t|^2$ and R=$|r|^2$, respectively) are visualized. 

\subsubsection{Spontaneous Emission}
The spontaneous emission dynamics of a two-level atom inside a waveguide  has been investigated in \cite{Law}. However, it is important to note that the presence of a linear dielectric medium with refractive index $n$ can cause changes for the lifetime of the two-level atom. Thus, we shall examine the decay rate of the two-level atom in a similar setup as in Fig. \ref{figmain}, which is initially prepared in the excited state
\begin{eqnarray}
\ket{\alpha}=\ket{0,+}.
\end{eqnarray}

One can project this state onto energy eigenstates given in (\ref{eq:eigenstate}) and then find the probability of the atom to remain in the excited state by taking into account the time evolution of the eigenstates as
\begin{eqnarray} \label{eq:spontane}
P(t)=\left| \frac{1}{2\pi} \int \diff k_n |e_k|^2 e^{-iE_k t}  \right|^2,
\end{eqnarray}
where, the negative values of $k_n$ indicate the scattering states incident from right. After straightforward algebra, we obtain the following expression for the probability of the two-level atom to remain in the excited state
\begin{equation} 
P(t)\simeq e^{-2 J_n t}. \label{eq:decay}
\end{equation} 

Comparing with the \cite{Law}, one can see that the spontaneous emission rate $\Gamma_0 =  2 V^2/v_g$ is simply replaced by $\Gamma_n = 2 V_n^2 n/v_g$ in the presence of the dielectric material with the refractive index $n$. As $V_n \geq V$, the presence of the dielectric enhances the radiative decay of the atom compared to the vacuum.

There is, however, another important prediction of this treatment, which was implied in \cite{fan, Law}. The spontaneous decay rate $\Gamma_{n}= 2 J_n$ is indeed equal to the half-power bandwidth of the transmission-reflection amplitudes. This then constitutes an indirect way of obtaining the spontaneous decay rate from transmission amplitudes.

\subsection{High-Q Bandreject Filter Application}
One can use the illustrative example above to build a high-Q bandreject filter as hinted in \cite{fan}. We can use the frequency of light $w_k$ to transfer signals where the central frequency of the filter created by the transmission line is $w=\Omega$, the cutoff frequencies are at $w_c = \Omega \pm J_n$, from which we obtain the bandwidth $\Delta w= 2 J_n$. 
From these filter characteristics, the quality factor becomes
\begin{eqnarray}
Q = \frac{w}{\Delta w} = \frac{\Omega}{2J_n}.
\end{eqnarray}

Here we note that the quality factor $Q$ is directly proportional to the level separation of the TLS, inversely proportional to coupling coefficient inside the dielectric. Any signal sent with a photon of frequency range $\Omega \pm J_n$ is reflected with more than $50\%$, which allows us to obtain a bandreject filter.

\begin{figure*}
\centering
\includegraphics[width=8cm]{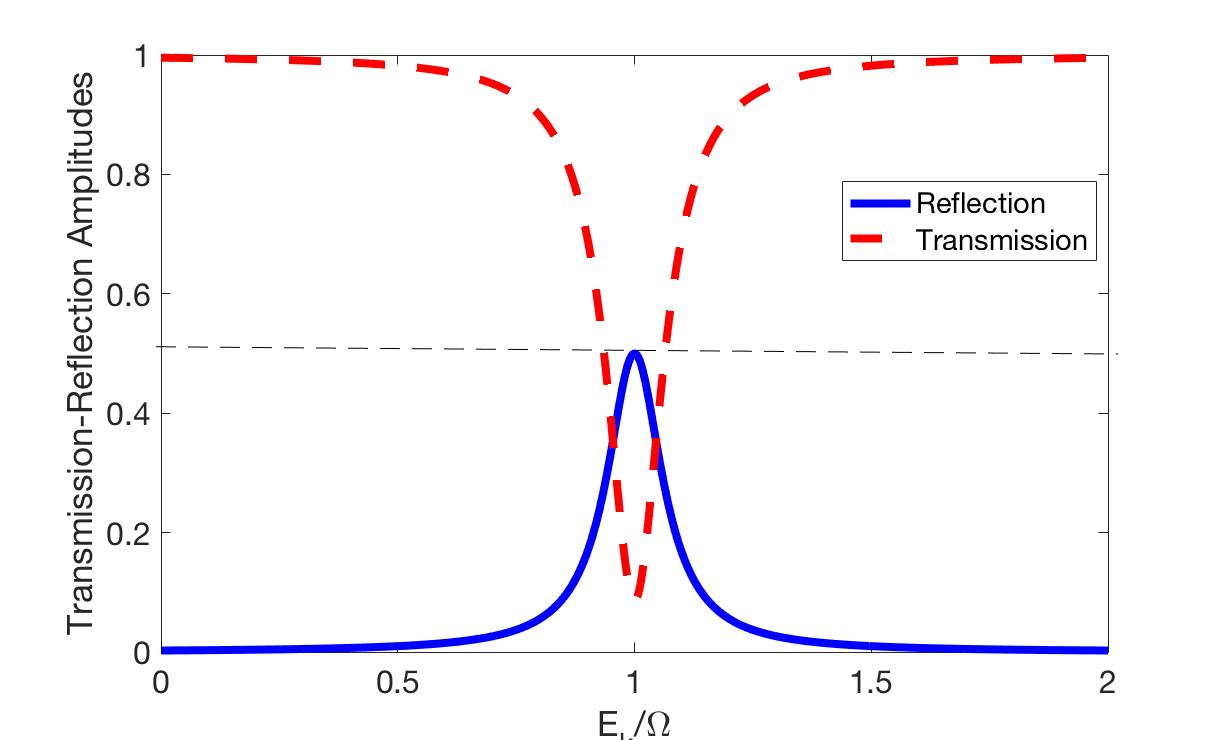}  
\includegraphics[width=8cm]{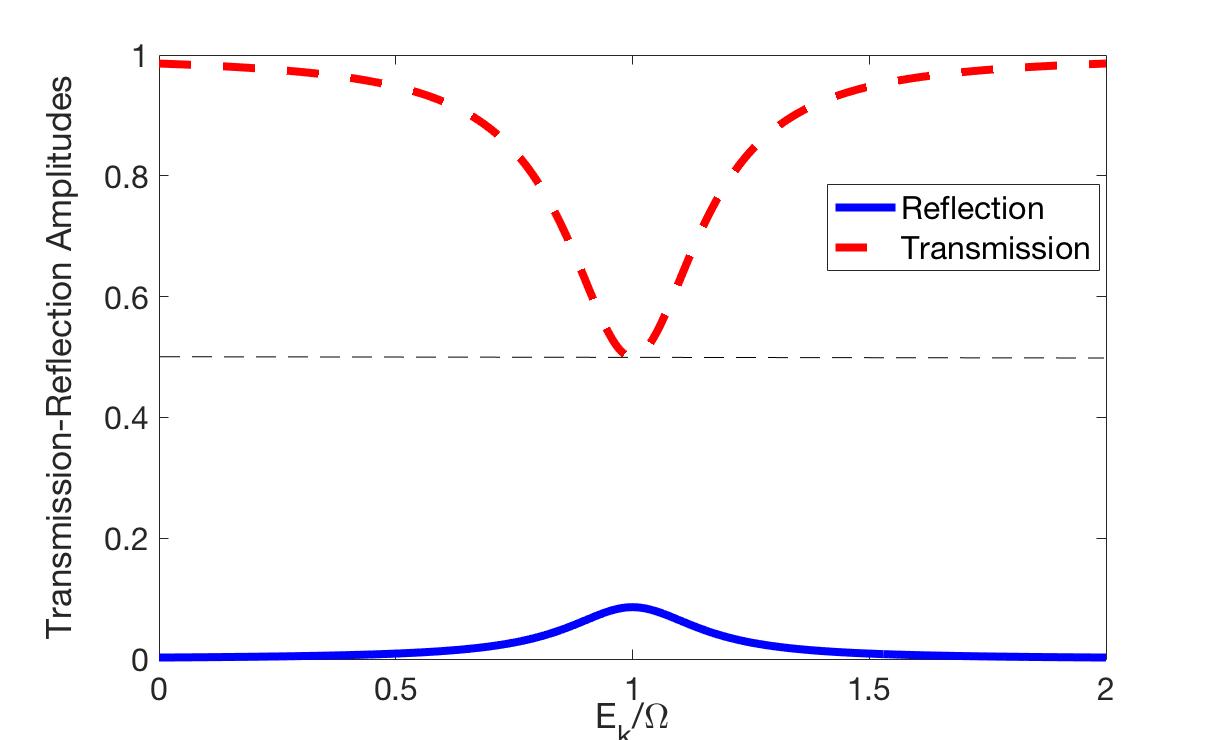} 
\caption{Transmission (dotted) and reflection (solide) amplitudes for $\gamma = \gamma_r$ (top) and $\gamma = \gamma_t$ (bottom) vs $E_k/\Omega$ in the presence of non-radiative dissipation with $J_n /\Omega = 0.05$.} \label{figdis}
\end{figure*}

\par\medskip\noindent
{\bf \emph{Further Remarks On Dissipation --}} 
Up to this point, we  assumed that the atom decays strictly radiatively and concluded that the spontaneous emission resulting from this radiative decay leads to a bandreject filter behaviour for the transmission amplitudes. 
Here, we discuss the behaviour of the two-level atom in the presence of a non-radiative decay rate $\gamma$. According to \cite{fandissipation}, the dissipation of the two-level atom can be taken into account by changing $\Omega \rightarrow \Omega- i \gamma /2$ in the S-matrix given in  (\ref{eq:smatrix}). We shall follow this procedure and obtain the transmission and reflection coefficients for a photon incident from far left as
\begin{subequations}  
\begin{align}
 r &= \frac{i \frac{\Gamma_n}{2}}{(\Omega - E_k ) - i \left(\frac{\Gamma_n}{2} + \frac{\gamma}{2} \right)}, \\
 t &= \frac{(\Omega - E_k) - i \frac{\gamma}{2}}{(\Omega - E_k ) - i \left(\frac{\Gamma_n}{2} + \frac{\gamma}{2} \right)},
\end{align}
\end{subequations}
where $\Gamma_n$ is the radiative decay rate in an optical medium with refractive index $n$ as found in equation (\ref{eq:decay}). From these transmission and reflection coefficients, $t$ and $r$ respectively, we obtain the corresponding amplitudes 
\begin{subequations}  
\begin{align}
 R=|r|^2 &= \frac{\Gamma_{n}^2}{(\Gamma_{n}+\gamma)^2+4(\Omega-E_k)^2},\\
  T=|t|^2 &= \frac{4(\Omega-E_k)^2+\gamma^2}{(\Gamma_{n}+\gamma)^2+4(\Omega-E_k)^2},
\end{align}
\end{subequations} 
where we can find that $T_{min}=\gamma^2/(\Gamma_{n}+\gamma)^2$ and $R_{max}=\Gamma_{n}^2/(\Gamma_{n}+\gamma)^2$. 
A quick analysis of these equations reveal that in the presence of  non-radiative decay rate $\gamma$, the reflection amplitude $R$ is modulated more dramatically than transmission amplitude $T$. In order to illustrate this, we define two critical non-radiative decay rates where reflection and transmission amplitudes are modulated heavily that they cannot overcome half-power barrier
\begin{eqnarray}
\gamma_r =  (\sqrt{2}-1) \Gamma_{n} \quad \text{and} \quad \gamma_t = (\sqrt{2}+1) \Gamma_{n}.
\end{eqnarray}

For any dissipation rate $\gamma>\gamma_{r/t}$, the reflection and transmission of light cannot be used as a signal filter, respectively. These two cases are illustrated in the Fig. \ref{figdis}.

The non-radiative dissipation of the two-level atom and the behavior of light in the optical boundaries presented under this formalism allow a direct comparison to be made with scattering experiments by using a quantum mechanical approach. It is also apparent from the Fig. \ref{figdis} using transmission amplitude is accurate means of sending signals as it is modulated relatively less by the presence of a non-radiative dissipation, even in cases where the non-radiative dissipation rate is higher than the radiative decay rate $\Gamma_n$.

\section{Fabry-P\'{e}rot Interferometer with a Two-Level Atom}
\begin{figure*}[!b]
\centering
\includegraphics[width=13cm]{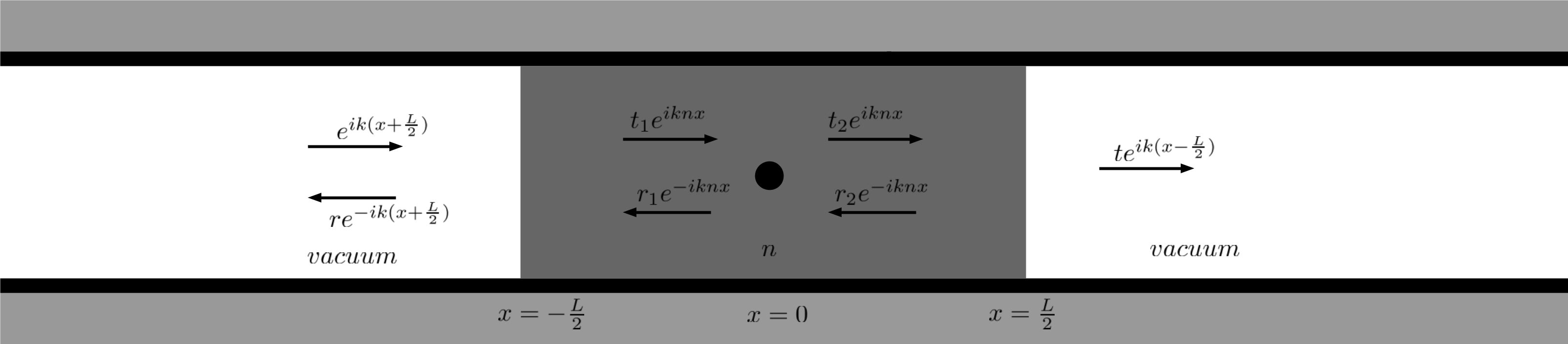}
\caption{Fabry-P\'{e}rot Interferometer with a TLS}\label{fig:fabry}
\end{figure*}

Finally, we illustrate an application of the present formalism via a modified Fabry-P\'{e}rot interferometer (MFPI) setup including a TLS as shown in Fig. \ref{fig:fabry}. The setup is very similar to a conventional Fabry-P\'{e}rot interferometer with the dielectric medium having width $L$; the only difference is that a TLS is situated halfway through the boundaries. We can define the following transfer matrices for different boundaries taken into account (\ref{eq:smatrix})

\begin{subequations}
\begin{align}
T_{TLS}&=
\begin{bmatrix}
1 - i \frac{J_n}{(\Omega - E_k)} & - i \frac{J_n}{(\Omega - E_k)} \\
 + i \frac{J_n}{(\Omega - E_k)} & 1 + i \frac{J_n}{(\Omega - E_k)}
\end{bmatrix},
\\
M_{1 \to n}&=
\begin{bmatrix}
\frac{n+1}{2}  & \frac{1-n}{2} \\
\frac{1-n}{2}   & \frac{n+1}{2}
\end{bmatrix}, \quad 
M_{n \to 1}=
\begin{bmatrix}
\frac{n+1}{2n}  & \frac{n-1}{2n} \\
\frac{n-1}{2n}   & \frac{n+1}{2n}
\end{bmatrix},\\
P_n &= 
\begin{bmatrix}
e^{-iknL/2}  & 0 \\
0   & e^{iknL/2} 
\end{bmatrix}.
\end{align}
\end{subequations}

\begin{figure*}[!b]
\centering
\includegraphics[width=9cm]{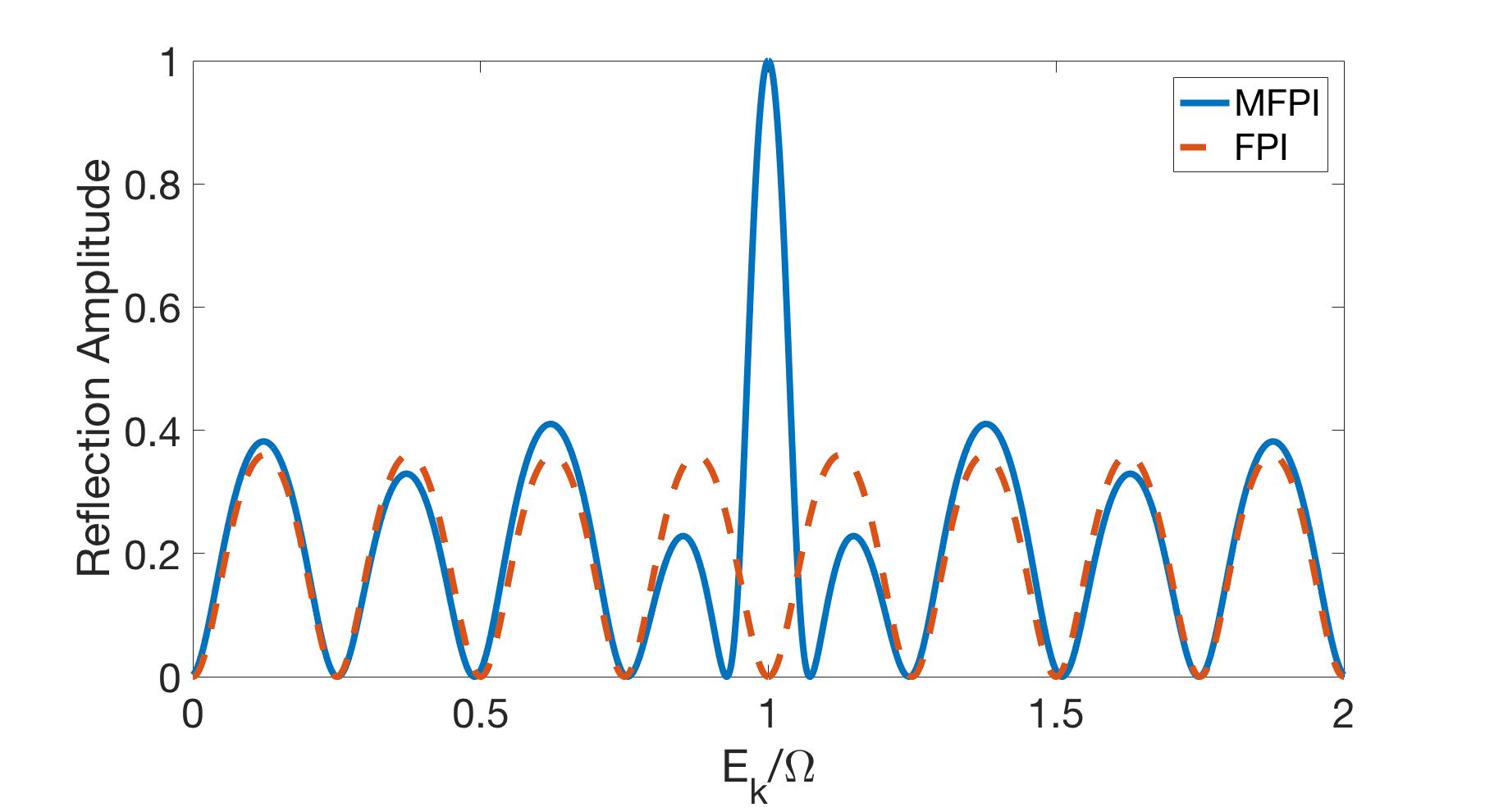}
\caption{The reflection amplitudes for MFPI and FPI. $J_n=0.025\Omega$ for MFPI and $J_n=0$ for FPI, $L = 2\pi /k_\Omega$, $n=2$, where $\hbar  v_g k_\Omega=\Omega$.} \label{fig:fabryref}
\end{figure*}

\noindent
where $T_{TLS}$ is the transfer matrix for the TLS, $M_{1\to n}$ is the transfer matrix governing the optical boundary from vacuum to the dielectric medium with the refractive index $n$ and $P_n$ is the propagation matrix. Regarding the MFPI as a black box, we obtain the following relation for the transmission and reflection coefficients
\begin{equation}
\begin{bmatrix}
1 \\
r
\end{bmatrix}
=
T_{box}
\begin{bmatrix}
t \\
0
\end{bmatrix},
\end{equation}
where we can identify $T_{box}=M_{1 \to n} \, P_n  \, T_{TLS} \,  P_n \, M_{n \to 1} $. After careful calculations, we can find the transport parameters and the excitation coefficient $e_k$ as
\begin{subequations}
\begin{align}
    r&= \frac{-i\Delta_k (e^{2ik_nL}-1)(n^2-1) - \Gamma_n e^{ik_nL} ((n^2+1)+(n^2-1)\cos(k_nL))}{-i\Delta_k ((n+1)^2-(n-1)^2 e^{2ik_nL}) + \frac{\Gamma_n}{2}((n+1)+(n-1)e^{ik_nL})^2 },\\
    t&=\frac{-4i \Delta_k n e^{ik_nL}}{-i\Delta_k ((n+1)^2-(n-1)^2 e^{2ik_nL}) + \frac{\Gamma_n}{2}((n+1)+(n-1)e^{ik_nL})^2}, \\
    e_k &=\frac{- 2 e^{\frac{1}{2}ik_nL} V_n}{e^{ik_nL}(n-1) (\Delta_k - i \frac{\Gamma_n}{2}) - (n+1) (\Delta_k + i \frac{\Gamma_n}{2}) },
\end{align}
\end{subequations}
where $\Delta_k=(E_k-\Omega)$. The reflection amplitude $R=|r|^2$ for FPI and MFPI are illustrated in Fig. \ref{fig:fabryref}. We note that for frequencies away from the resonance frequency $\Omega$ the MFPI behaves similar to FPI. However, there is a remarkable difference when the photon is incident in resonance. In that case, one observes the quantum mirror behavior of the TLS dominating over the FPI.

The spontaneous emission characteristics of the MFPI differs significantly from that of the TLS discussed above. In TLS case, the existence of the dielectric was completely lumped into the coupling energy $J_n$. In this case, however, we cannot lump the effect of the dielectric medium and should investigate the spontaneous emission characteristics thoroughly by a modified version of equation (\ref{eq:spontane}):
\begin{eqnarray} \label{eq:int}
P(t)= A \left| \frac{1}{2\pi} \int \diff (k_n) |e_k|^2 e^{-iE_k t}  \right|^2,
\end{eqnarray}
where the parameter $A$ ensures the normalization which is no longer preserved in equation (\ref{eq:spontane}) as the normalization of the vacuum states in different dielectric media varies. Nonetheless, the physics of the spontaneous emission is encoded in the denominator inside the integral and therefore unaffected by the normalization. Note that as $L \to 0$ or $n \to 1$, the denominator of the expression inside this integral takes the expected form of a TLS situated inside vacuum, whereas the normalization is not recovered for the former case. Finally, we note that in the weak-coupling regime ($J_n << \Omega$), $|e_k|^2$ is highly peaked around the resonance and therefore we can linearize the phase factors as $e^{ik_nL} \simeq e^{i k_{\Omega_n} L}$, where $k_{\Omega_n}$ is the wave-vector inside the dielectric medium with refractive index $n$ corresponding to the resonance energy of the atom $\Omega$. The poles of $|e_k|^2$ in (\ref{eq:int}) can be found as
\begin{equation}
    p_{1/2}\simeq \frac{\Gamma_n}{2} \frac{\left(n^2-1\right) \sin (k_{\Omega_n}L) \pm i 2 n}{n^2+1- (n^2-1) \cos (k_{\Omega_n}L)}. 
\end{equation}

The decay rate can be found through Contour integration to be $\Gamma = 2 |Im[p_{1/2}]| $, which is
\begin{eqnarray}
\Gamma \simeq \Gamma_n \frac{2 n}{n^2+1- (n^2-1) \cos (k_{\Omega_n}L)} \label{eq:gamma}.
\end{eqnarray}
We can show that for $n=1$, $\Gamma = \Gamma_0$ corresponding to the vacuum and as $L \to \infty$, averaging over a period of $\cos(k_{\Omega_n} L)$, $\Gamma=\Gamma_n$ as expected.

Choosing $L$ such that it corresponds to the resonance wavelength $\lambda_{\Omega_n}$ (e.g. $L = m \lambda_{\Omega_n}, m=1,2,..$) the decay rate is highest with $\Gamma= n \Gamma_n$.  The intuitive explanation is as follows: A photon which is emitted from the TLS travels a distance $\lambda_{\Omega_n}$, gets partially reflected by the optical boundary comes back to the position of TLS. This photon then constructively interferes with itself, hence increasing the probability that the atom decays.

Conversely, choosing $L$ such that $L = (m+1/2) \lambda_{\Omega_n}$ ($m=1,2,..$) we obtain the lowest decay rate, i.e. highest life-time, with $\Gamma= \frac{\Gamma_n}{n}$. In this case, the emitted photon comes back and interferes destructively with itself, decreasing the probability of emission. Therefore, the radiative decay of the TLS can be tuned through a MFPI between $n \Gamma_n$ and $\Gamma_n/n$ by choosing the length $L$ correspondingly.

\section*{Further Remarks on Waveguide and Cavity QED}

For the scope of this paper, we primarily focus on the effects of changing dielectric media inside a waveguide while linearizing the displacement field and the coupling $V_n^2$ between the TLS and photons. Realistically, these quantities are highly dependent on the interactions between the dipoles of dielectric media and photons, hence on the frequency of the incoming photon. The nature of $V_n$ and how it relates to $V$ is an important research question that needs to be addressed in future work, which includes a similar treatment to the one in Chapter 7 from \cite{dutra2005cavity}. 

Even in its highly linearized current form, our treatment proposes a link between the waveguide and cavity QED, where a Fabry-Perot interferometer is introduced as a means of tuning the coupling $V_n^2$ between the TLS and the photon. By changing the decay rate $\Gamma_n$, which changes with respect to $L$ as found in (\ref{eq:gamma}), one practically changes the effective coupling $ V_n^2 $ between the TLS and the incoming photon. This is only one example of utilizing cavities as a means of tuning the coupling strength inside a waveguide, hence illustrating an important link between waveguide and cavity QED.

\section{Conclusion}
In this paper, we have presented a quantum mechanical formalism including the real-space Hamiltonian of a two-level atom coupled to a photon in a continuum that allows treatment of systems with optical medium inside the continuum. We show that the results obtained by using this formalism is inline with the ones presented in \cite{fan,Law} for corresponding limiting cases. In this pursuit, we discuss the illustrative example of photon interacting with a two-level atom in the presence of optical medium. The generalized formalism we propose here allows removing any assumptions regarding the discontinuity of the continuum caused by the delta-scattering potential. This enables us to investigate photon scattering in arbitrary position space potential $V(x)$. Furthermore, the modified Fabry-P\'{e}rot study shows that changing the length of the interferometer has significant implications for the lifetime of the TLS; where adjusting lifetime can be useful for switching and quantum computing application. Our calculations reveal that using a Fabry-Perot interferometer allows one to tune the radiative decay rate of a TLS between $ n \Gamma_n$ and $\Gamma_n/n$.

\par As a part of our future work, we aim to explore the effect of different scattering potentials which  allows further generalization of the scope of systems studied using such formalism, and provide basis for comparison including wide range of experimental results.  We also aim to incorporate the interaction of multiple-photon transport as well as wave-packets with the two-level atom and compare our formalism against the existing literature \cite{kocabas,kocabas2,shi}. Expanding the theoretical framework here to various 1-D geometries is also intended for future applications as presented in \cite{rephaeli}. Furthermore, our formalism can find further applications in resonator, cavity and multi-atom systems by expanding the findings of \cite{ilkehoca,cavity,Law} to include changes in dielectric medium.



%

\section*{Acknowledgements}
The authors would like thank Professor Atac Imamoglu for intellectually stimulating discussions, Professor Teoman Turgut for insightful  comments on the manuscript and Professor \c{S}\"{u}kr\"{u} Ekin Kocaba\c{s} for inspiration and guidance on the modified Fabry-P\'{e}rot interferometer study.


\appendix

\section{Proof of Concept}\label{proofofconcept}

A scattering eigenstate $\ket{E_k}$, for the Hamiltonian in (1), is given by the equation: 
\begin{equation} 
\begin{split}
	\ket{E_k}=\int_{-\infty}^{\infty} \diff x &\: \Bigg( e^{ikx} u_R (x) + e^{-ikx} u_L (x)  \Bigg) \phi^\dag(x) \ket{0,-} \\ &+ e_k  \ket{0,+},
\end{split}
\end{equation} 
where we denote $e_k$ as the probability amplitude for the excitation of the atom, and $u_R(x)$ and  $u_L(x)$ are the complex amplitudes of the right and left moving photons, respectively.

For a photon incident far form the left, $u_R(x)$ and  $u_L(x)$ are defined as follows
\begin{subequations}
\begin{align}
\lim_{x \to \infty} u_R(x) = t, \\
\lim_{x \to \infty} u_L(x) = 0, \\
\lim_{x \to -\infty} u_R(x) = 1, \\
\lim_{x \to -\infty} u_L(x) = r,
\end{align}
\end{subequations} 
where $r$ and $t$ are the reflection and transmission coefficients, respectively. It is important to note that, by ansatz, the Fourier coefficient $\bar u_R(k')$ ($\bar u_L(k')$) is zero for any $k'<-k$ ($k'>k$) to divide strictly right-moving and left-moving part of the eigenstate. It is also important to note that the modulations in $u_R(x)$ and $u_L(x)$ happen between $0^-$ and $0^+$ due to the delta-scattering potential $V \delta(x)$.

Solving the eigenvalue equation $\hat H \ket{E_k}=E_k \ket{E_k}$ and applying the slowly varying field approximation, one obtains the set of equations
\begin{subequations}
\begin{align}  
		i \hbar v_g (1-t-r) + 2 e_k V' &= 0, \label{eq:bounda} \\
		i \hbar v_g (1+r - t) &=0, \label{eq:boundb} \\
		V't + (\Omega -E_k) e_k&=0.
\end{align} 
\end{subequations}
We can solve them to obtain
\begin{equation}  \label{eq:9}
	t= \cos b e^{ib},\quad r=i\sin b e^{ib}, \quad e_k = -\frac{ v_g}{V'}\sin b e^{ib},
\end{equation} 
where the phase shift is given by $b= \arctan\left(\frac{V'^2}{ v_g(\Omega-E_k)} \right)$, which allows us to redrive the findings of \cite{fan,Law} using our formalism.

\section{Sanity Checks for The Final Result} \label{ap2}
In this appendix, two sanity checks will be performed to show the consistency of (\ref{experiment}). The first check is for the case when there is no dielectric material present. For this case, (\ref{experiment}) becomes
\begin{subequations} 
\begin{align}
 r &= i \sin b  e^{ib},\\
  t &= \cos b e^{ib}, \\
     e_k&=-\frac{v_g}{V} \sin b e^{ib},
\end{align}
\end{subequations}
where the phase shift is $b=\arctan\{ V^2/[v_g(\Omega-E_k)]\}$, hence resulting in the (\ref{eq:9}).

The final sanity check is regarding the conservation of probability current, or classically energy. Since there is no non-radiative decay taken into account in these calculations, the sum of transmission and reflection amplitudes is conserved such that
\begin{equation}
|r|^2+|t|^2 = 1.
\end{equation} 
Note that this can be used as evidence for conservation of energy.

\end{document}